# Emergence of superconductivity in the cuprates via a universal percolation process

Damjan Pelc,[1,†,#] Marija Vučković,[1,†] Mihael S. Grbić,[1] Miroslav Požek,[1,*] Guichuan Yu,[2] Takao Sasagawa,[3] Martin Greven,[2,*] Neven Barišić[2,4,*]

[1]Department of Physics, Faculty of Science, University of Zagreb, Bijenička 32, HR-10000, Zagreb, Croatia

[2]School of Physics and Astronomy, University of Minnesota, Minneapolis, MN 55455, USA

[3]Materials and Structures Laboratory, Tokyo Institute of Technology, Kanagawa 226-8503, Japan

[4]Institute of Solid State Physics, TU Wien, 1040 Vienna, Austria

†These authors have contributed equally

#Present address: School of Physics and Astronomy, University of Minnesota, Minneapolis, MN 55455, USA

*Correspondence to: mpozek@phy.hr, greven@umn.edu, neven.barisic@tuwien.ac.at

A pivotal step toward understanding unconventional superconductors would be to decipher how superconductivity emerges from the unusual "normal state" upon cooling. In the cuprates, traces of superconducting pairing appear above the macroscopic transition temperature $T_c$, yet extensive investigation has led to disparate conclusions. The main difficulty has been the separation of superconducting contributions from complex normal-state behaviour. Here we avoid this problem by measuring the nonlinear conductivity, an observable that is zero in the normal state. We uncover for several representative cuprates that the nonlinear conductivity vanishes exponentially above $T_c$, both with temperature and magnetic field, and exhibits temperature-scaling characterized by a nearly universal scale $T_0$. Attempts to model the response with the frequently evoked Ginzburg-Landau theory are unsuccessful. Instead, our findings are captured by a simple percolation model that can also explain other properties of the cuprates. We thus resolve a long-standing conundrum by showing that the emergence of superconductivity in the cuprates is dominated by their inherent inhomogeneity.

Despite tremendous experimental and theoretical efforts over the past thirty years, the nature of the superconducting fluctuation regime of the cuprates remains intensely debated[1]. Experimentally, the problem has been approached using bulk probes, such as transport[2-8], magnetic susceptibility[9,10] and optical conductivity[11]; surface sensitive probes[12]; and local probes, such as muon spin rotation[13]. Some studies point to the possible persistence of superconducting pairing well above $T_c$, which has been taken as an indication of pre-formed Cooper pairs related to the appearance of the pseudogap[11,14]. Other studies indicate that traces of superconductivity emerge at somewhat lower temperatures and are most prominent at moderate doping[6,7,9]. High pairing onset temperatures have been related to exotic normal-state physics[15,16] and unconventional pre-pairing[17,18], with profound consequences for the mechanism of cuprate superconductivity. However, terahertz and microwave conductivity[2,3,8] as well as magnetometry experiments[10] consistently detect superconducting contributions only near $T_c$, irrespective of doping. The resolution of this puzzle would be a crucial step toward understanding the high-$T_c$ cuprates.

Previous experiments suffer from a crucial problem: the inability to reliably establish the non-superconducting normal-state contribution in order to extract a superconducting signal. Typical approaches involve the extrapolation of high-temperature behaviour or the suppression of superconductivity by a magnetic field. The situation is further convoluted due to the complexity of the cuprate phase diagram, which features a doping-dependent pseudogap, as well as universal and compound-specific ordering tendencies that manifest themselves differently in different experimental observables[1]. The presence of different kinds of disorder in these complex oxides poses yet another complication[19]. Data are often analyzed within the Ginzburg-Landau (GL) framework, with corrections to the original mean-field theory[4,5,8], yet this has not resulted in a unifying picture.

The absence of any discernible signal due to non-superconducting contributions makes nonlinear conductivity uniquely suitable to study and model superconductivity emergence. We apply this probe to a number of cuprate families and a variety of experimental conditions. The measurements unambiguously show that superconductivity emergence is *not* controlled by $T_c$, the crucial qualitative feature of GL theory, but rather by a universal scale $T_0$ that is nearly independent of compound and doping (in the studied doping range $p$ = 0.08 - 0.19). Thus, we show that a comprehensive description of the data is not possible with GL-type approaches. We then use a simple model to explain the data: the superconducting gap is known to be spatially inhomogeneous, which results in a distribution of local transition temperatures and naturally leads to percolation. Percolation, and the scale-free fractal structures that emerge from it, is a well-known and ubiquitous phenomenon: first investigated in the context of polymer growth, it has since been formulated as a mathematical concept and applied to systems as diverse as random resistor networks, organic molecular gels, dilute magnets, the spread of diseases, and the large-scale structure of the universe[20,21]. The basic ingredient in percolation theory is inhomogeneity; we find that evoking inhomogeneity is essential to understand superconductivity emergence in the cuprates. Remarkably, the minimal percolation model is sufficient to capture the unusual exponential temperature- and magnetic-field dependences of the nonlinear conductivity that we observe. We also report complementary linear conductivity measurements and take a fresh look at prior experimental results (torque magnetometry[10], resistivity[4], Seebeck coefficient[5] and tomographic density of states[22]), to demonstrate that the emergence of superconductivity can be consistently explained with this minimal model.

**Nonlinear response.** Nonlinear planar response, for current-flow along the CuO$_2$ planes, is measured with a sensitive contact-free method[23] (see Methods). The response can be analyzed by decomposing the signal into harmonics,

$$J = \sigma_1 K + \sigma_2 K^2 + \sigma_3 K^3 + \dots ,$$

where $J$ is the response of the sample to an external field $K$ (electric or magnetic), $\sigma_1$ the linear response tensor, and $\sigma_2$, $\sigma_3$, etc., the correction nonlinear tensors. Here we discuss the third harmonic $\sigma_3$, the lowest-order conventional correction to the linear response (the second-harmonic $\sigma_2$ can only appear if time reversal or inversion symmetry is broken[24] and is not discussed here). In any alternating-field experiment, magnetic and electric fields are related, and therefore it is arbitrary if one designates the signal at frequency $3\omega$ as proportional to nonlinear conductivity or (complex) susceptibility. Complementary linear conductivity measurements are performed with a microwave cavity perturbation technique[8] (see Methods).

**Temperature dependence.** Measurements of the in-plane linear and nonlinear response were performed for three representative cuprate families: a nearly optimally-doped sample of HgBa$_2$CuO$_{4+\delta}$ (Hg1201), which in many respects can be regarded a model cuprate[25,26]; an optimally-doped YBa$_2$Cu$_3$O$_{7-\delta}$ (YBCO) sample with 3% of Cu substituted by Zn (YBCO-Zn), where Zn dramatically affects the superconducting properties[27]; and La$_{2-x}$Sr$_x$CuO$_4$ (LSCO), spanning a wide range of doping across the superconducting dome (see Table 1). For all samples, $\sigma_3$ exhibits qualitatively the same temperature dependence (Fig. 1a and Supplementary Fig. 1): no signal at high temperatures, a peak at a temperature that we designate as $T_c$, consistent with previous work[8] (see Supplementary Notes), and a step-like feature below $T_c$. The signal magnitude depends on sample size and shape, and thus is normalized to the peak value. We note that $T_c$ as obtained from $\sigma_3$ agrees with the temperature of the peak in the real part of the linear microwave conductivity, which in turn corresponds to the value determined from magnetic susceptibility measurements[8].

Previous investigations suggested agreement between experiments and GL theory (with various modifications to the theory[4,5,8]) for particular cuprate compounds at particular doping levels. In line with these investigations, we have attempted to analyze our results within the GL framework. Within this framework, we would expect an approximately power-law temperature dependence of $\sigma_3$ (see Ref. [28] and Supplementary Notes), and a scaling of data for different compounds with the characteristic scale $T_c$. Figure 1b shows our nonlinear conductivity data in dependence on the GL reduced temperature $\ln(T/T_c)$ compared to a calculation of $\sigma_3$ using anisotropic GL theory beyond mean field (see Supplementary Notes), similar to Ref. [8]. The theory predicts a temperature dependence of $\sigma_3$ that is clearly incompatible with experiment; the agreement cannot be improved by any tuning of the parameters, such as a different definition of $T_c$ (see Supplementary Fig. 2). Even more importantly, the expected scaling is absent: $T_c$ is not the characteristic temperature scale for superconductivity emergence. The scaling argument is valid regardless of the manner in which GL theory is modified. However, the data are remarkably similar on an *absolute* temperature scale: a simple shift by a sample-dependent temperature $T_\pi$ (that is slightly larger than $T_c$) leads to the remarkable data collapse shown in Fig. 1c. This implies that a mechanism that gives rise to approximately exponential behaviour with a single temperature scale $T_0$ underlies the emergence of superconductivity. A similar exponential dependence can also be deduced from linear conductivity (inset in Fig. 1c) and torque magnetometry[10] experiments, indicating its robustness. Clearly a framework other than GL is needed to explain the data.

Since nanoscale electronic inhomogeneity is well documented in the cuprates, e.g., from nuclear magnetic resonance[29-31] and scanning tunneling microscopy (STM)[12,32] measurements, we now attempt to gain understanding through a simple percolation model. The basic ingredient of the model is spatial inhomogeneity of local superconducting gaps, with a distribution width that is characterized by the scale $k_B T_0$. This distribution corresponds to superconducting patches that proliferate upon cooling, and macroscopic superconductivity then emergs via a percolative process[33,34]. We calculate the response assuming nearest-neighbor site percolation, although the result does not critically depend on the details of the scenario (see Supplementary Notes). For simplicity, we take the material to be made of perfectly connected square or cubic patches that are either non-superconducting, with a normal resistance $R_n$, or superconducting, with a nonlinear resistance $R_s$ that depends on the current through the patch (see Supplementary Notes). The fraction $P$ of superconducting patches depends on temperature: $P \rightarrow 0$ at high temperatures and $P \rightarrow 1$ well below $T_c$. At the critical concentration $P_\pi$, the system percolates – a connected, sample-spanning superconducting cluster is formed. $P_\pi$ only depends on the dimensionality of the system[20] and on the details of the percolation scenario (e.g., site versus bond percolation), and it corresponds to the temperature $T_\pi$ that can be viewed as the "true" underlying resistive $T_c$ in the limit of small currents. In principle, the full temperature dependence of $P$ can be obtained from the underlying gap distribution, but the distribution must be known (or assumed). However, to lowest order, any reasonable distribution yields a linear dependence of $P$ on temperature close to $P_\pi$ (e.g., see Supplementary Fig. 3). We therefore approximate $P_\pi - P = (T - T_\pi)/T_0$. The temperature-dependent linear and nonlinear responses are then obtained via an effective medium calculation (see Supplementary Notes), which yields functions that decay nearly exponentially, in very good agreement with the experimental $\sigma_3$ and $\sigma_1$ (Fig. 1c). Within the nearest-neighbor site percolation model, two values of $P_\pi$ are possible: $\approx 0.3$ for three-dimensional and $\approx 0.6$ for two-dimensional percolation. Better agreement is obtained with $P_\pi \approx 0.3$ (see Supplementary Fig. 4), which suggests essentially three-dimensional superconductivity emergence[20] in the samples studied here. We note that we study the in-plane response, and thus the only role of inter-plane coupling in the percolation model is to determine the effective dimensionality, and hence the percolation threshold.

For $P_\pi \approx 0.3$, the resultant characteristic temperature $T_0$ lies in a narrow range for all investigated samples (see Table 1), $T_0 = 27 \pm 2$ K, and hence is de facto universal. If we assume 2D percolation and $P_\pi \approx 0.6$, the agreement between $\sigma_3$ and $\sigma_1$ is not as good, and the corresponding $T_0$ is smaller by about a factor of two. We emphasize that the calculated $\sigma_3$ is effectively insensitive to model details such as the parameters of the patch nonlinear response $R_s$, rendering $T_0$ the sole parameter (within a given percolation model). This insensitivity to specifics is a consequence of percolation physics, where model details are unimportant close to the threshold and the response of the largest clusters dominates.

An important feature can be inferred from the comparison of linear and nonlinear response. Within the effective medium calculation, the linear conductivity determines the net current through the sample, given an applied electric field. Yet the nonlinear resistance of the superconducting patches, $R_s$, is current-dependent. The third-harmonic signal therefore depends on the third power of the current (to lowest order), which implies that $\sigma_3 \propto \sigma_1^3$. This is indeed borne out by experiment, as seen in Fig. 1d. In contrast, in GL theory both responses are determined by the electric field, and their ratio has a more complex temperature dependence (see Ref. [28] and Supplementary Notes). The *apparent* characteristic temperature scales for $\sigma_3$ and $\sigma_1$ therefore differ because of the nonlinear nature of $\sigma_3$, but the underlying scale $T_0$, which determines the range of superconducting pairing emergence, is the same for both responses. This also implies that the superconducting contribution to the linear response should be discernable up to significantly higher temperatures than the nonlinear part, if the experimental signal-to-noise ratios are similar. Measurements throughout the phase diagram of LSCO consistently confirm this trend (Fig. 1d), which strongly supports the percolation model.

**Magnetic-field effect.** As a further test of the model, we investigate the influence of an external magnetic field on the emergence regime. Although the quantitative effects of a magnetic field are difficult to determine within our simple effective-medium approach, we can make qualitative predictions. One would expect the field to greatly influence the superconducting percolation process. Both the critical current of a superconducting patch and the number of patches decrease with increasing field. Above a characteristic field $H_0$, the critical currents of all patches are small, except for the sample-spanning cluster at temperatures below $T_\pi$. At fields significantly above $H_0$, the nonlinear response should therefore only exhibit a step-like feature close to $T_\pi$. Furthermore, $H_0$ is related to the macroscopic critical field $H_{c2}$, as both fields are determined by the underlying superfluid stiffness: $H_0$ describes the properties of the finite-sized clusters below and above $T_\pi$, whereas $H_{c2}$ pertains to the sample-spanning cluster below $T_\pi$. In agreement with these expectations, we find that an external magnetic field strongly suppresses the nonlinear response, rendering it step-shaped far above $H_0$ (Fig. 2a). Once the high-field step-like response is subtracted (see Supplementary Notes and Supplementary Fig. 5), the data for all samples exhibit universal scaling (Fig. 2b). We apply the same effective medium calculation as for the temperature dependence, assuming a phenomenological power-law dependence of the effective patch critical current on $H/H_0$ (see Supplementary Notes and Supplementary Fig. 6), and find good agreement with experiment (Fig. 2b). For $H >> H_0$, only large superconducting clusters survive. Since the cluster-size distribution in any percolation model is generally exponential for the largest clusters[20], this leads to an exponential-like field dependence of the high-field response, as also observed in prior torque measurements[10].

$H_0$ is about two orders of magnitude smaller than $H_{c2}$, consistent with the percolation scenario, since $H_0$ is a property of the *average* (small) cluster. As seen from Fig. 2b, the doping dependencies of the two characteristic fields are remarkably similar, including a minimum close to the "1/8 anomaly" of LSCO and YBCO[5,35,36]. The substitution of 3% Cu with Zn in YBCO causes a dramatic decrease of $H_0$, in agreement with established effects of Zn on superconductivity in cuprates[27].

**Discussion.** The percolation picture is in excellent agreement with the temperature and magnetic field dependencies of $\sigma_1$ and $\sigma_3$, and one would expect it to provide an explanation of other experimental results. As demonstrated in Fig. 3, this is indeed the case. Torque magnetometry measurements[10] exhibit both an exponential signal decrease above $T_c$ (Fig. 3d) and a universal temperature scale $T_{0,\text{torque}}$. An exponential tail is observed in the dc conductivity[4] of YBCO (Fig. 3b) and the Nernst effect[5] of Eu-LSCO (Fig. 3c). Although the latter result can be described by 2D Gaussian theory close to $T_c$, where corrections to the simple percolation picture are expected, once those data are plotted on an absolute temperature scale, the exponential tail is apparent and reveals the same underlying temperature/energy scale $T_0$. The exponential dependences at temperatures well above $T_c$ are a consequence of the tail of the superconducting gap distribution, and for $\sigma_1$ and $\sigma_3$ the effective medium calculation smoothly continues this dependence down to $T_c$. We note that the percolation model discussed here is somewhat different from the standard textbook case, in that both the normal and percolating (superconducting) patches have nonzero conductivity. Therefore, instead of a discontinuity at $T_\pi$ and power-law behaviour above the percolation temperature (that is predicted if one of the phases is insulating[20]), the calculation yields smooth exponential-like behaviour. Yet the underlying distribution of superconducting cluster sizes should still be scale-free (i.e., follow a power law). A signature of this might be observed with other experimental probes, e.g., recent optical pump-probe experiments uncovered hitherto unexplained power-law superconducting correlations[37] above $T_c$.

We emphasize that the heterogeneity that gives rise to superconducting percolation is qualitatively different from the disorder discussed previously in the context of 'dirty' and granular superconductors[38-40]. In alloys[38] and films[39,40], the electronic mean-free path is extremely shortened by scatterers, while in granular materials differing Josephson couplings between granules cause superconducting percolation[38]. Yet here we find that nanoscale *gap* inhomogeneity is crucial: the superconducting gap, and hence the local $T_c$ displays spatial variations and causes the percolation we observe. Related gap disorder (on scales much larger than the superconducting coherence length) has been employed previously in modeling the magnetization of select cuprate and other superconductors[41], but not applied universally or used to calculate transport properties. Inhomogeneity and a residual zero-temperature component of unconsidered carriers has been shown to be essential to understand low-temperature superfluid density and optical response of different cuprates[42]. Spatial gap inhomogeneity also naturally explains the 'gap filling' recently observed in a tomographic-density-of-states photoemission experiment[22]. As demonstrated in Fig. 3f&g, a quantitative description of this result can be obtained simply by positing that the measured density of states is an average over spatial regions with inhomogeneous gaps, again with a distribution width of $k_B T_0 \sim 3$ meV, which further supports the percolation scenario (see Supplementary Notes for details).

Perhaps the most unexpected result of our study, which covers the doping range from the very underdoped ($p = 0.08$) to the overdoped ($p = 0.19$) part of the phase diagram, is the existence of a (nearly) doping- and sample-independent percolation scale $T_0$, which implies a common origin of the gap disorder in all cuprates. Doping does not significantly alter this scale, but affects the macroscopic $T_c$ or, equivalently, the critical percolation temperature. Several distinct types of disorder are generally present in the cuprates: the lamellar structure is intrinsically frustrated, leading to structural inhomogeneity; the hole doping process necessarily introduces defects into the crystal structure; and doping a strongly correlated electronic system may induce electronic frustration and inhomogeneity. These different kinds of disorder typically are compound- and doping-dependent[19,43], and various experimental techniques have been used to study them. Residual resistivity, a measure of disorder, is compound-dependent, and can be very small in compounds such as Hg1201[25,26]. Furthermore, quantum oscillation experiments point to a considerable degree of doping (hole concentration) homogeneity in some oxygen-doped compounds[44,45]. However, this does not preclude considerable nanoscale electronic inhomogeneity. The cuprates also appear to exhibit inherent structural inhomogeneity, an elegant demonstration of which comes from conductivity and hydrostatic relaxation experiments that show stretched exponential behaviour characteristic of glassy materials[43]. Moreover, X-ray experiments find complex fractal interstitial-oxygen-dopant structures linked to percolative superconductivity[46]. Local electrostatic disorder has been studied via nuclear quadrupole resonance and revealed that LSCO exhibits higher levels of such disorder[29] than oxygen-doped cuprates[30,31] such as YBCO and Hg1201. However, importantly, none of these experiments directly detect superconducting gap disorder, making it difficult to establish a relationship between electrostatic/doping inhomogeneities and superconducting gap distributions. STM does probe local gap distributions on the sample surface, but has been applied only to a select number of cuprates, and it is not trivial to separate the superconducting gap from the more inhomogeneous higher-energy (pseudo)gap[12]. We emphasize that the gap distribution (with width $k_B T_0$) relevant for our model likely is not precisely the same as the gap distribution seen by STM, but rather a coarse-grained distribution of mean local gaps (averaged over the local superconducting coherence lengths). Nevertheless, STM clearly reveals disorder structures in both underdoped and overdoped[32] $Bi_2Sr_2CaCu_2O_{8+\delta}$, and extended analysis shows a correlation between the presence of inhomogeneous high-energy gaps and superconductivity[47]. A recent phenomenological model[48] based on inhomogeneous, temperature- and doping-dependent (de)localization of one hole per unit cell explains the main features of the cuprate phase diagram and superconductivity. It is conceivable that a universal scale $k_B T_0$ emerges via a complex renormalization of these high-energy localization gaps[48] (see also Supplementary Notes). In this case, the gap disorder would not necessarily be related to any local doping inhomogeneity: the material may be homogeneously doped, yet possess an underlying gap distribution.

The existence of exponential behaviour with a universal scale $T_0$ also shows that expected GL superconducting fluctuations are considerably weaker than inhomogeneity effects. Conversely, if GL fluctuations were important, the simple percolation model with a single $T_0$ would not describe the measurements. This furthermore points to three-dimensional percolation – except perhaps in the special case of strong stripe correlations in La-based cuprates[5,49] such as $La_{1.875}Ba_{0.125}CuO_4$ and $La_{1.8-x}Eu_{0.2}Sr_xCuO_4$ – since the strong GL fluctuations expected in the two-dimensional case should significantly broaden the onset of superconductivity.

The unexpected scaling of nonlinear and linear conductivity for widely different cuprates is a benchmark for any theory of superconducting pre-pairing in these materials. Taking into account the well-established fact that significant gap inhomogeneity is present in the cuprates, we have provided a simple framework for understanding the experiments in which disorder plays a pivotal role. Our results thus show that intrinsic and universal superconducting gap disorder is highly relevant to understanding the superconducting properties of the cuprates.

**Materials and Methods**

*Samples.* The Hg1201 and LSCO samples are single crystals of well-established quality used in previous work[50,51] with volumes of about 1 mm³. YBCO-Zn is an oriented powder sample with 3% Cu substituted with Zn, which enables us to discern effects of intentionally-introduced $CuO_2$ plane disorder. This sample was prepared using a standard solid-state reaction, used in prior Zn nuclear quadrupole resonance experiments and characterized in detail[27]. See Table 1 for additional sample information.

*Linear and nonlinear conductivity.* The nonlinear conductivity experiments were performed with a contactless radio-frequency two-coil setup, with excitation frequency $\omega/2\pi$ = 17 MHz and phase sensitive detection at $3\omega/2\pi$. This configuration enables high sensitivity and eliminates Joule heating nonlinearities. The coil system is kept at a constant temperature of the liquid helium bath, while the sample temperature is varied independently. A thin-walled glass tube separated the vacuum of the sample space from the liquid helium bath and introduced no distortions to the signal. The sample was mounted on a sapphire holder with temperature control sensitivity better than 1 mK. The setup was previously tested under various conditions[23,52] (see also Supplementary Methods). The electric fields within the samples may be estimated using Maxwell's equations, which gives an amplitude $E \sim B\omega L$, where $B$ is the magnetic field amplitude, $\omega \sim 2\pi \cdot 20$ MHz the oscillation frequency, and $L \sim 2$ mm the typical linear sample dimension. The amplitude of the magnetic field was deliberately kept small, and estimated to be about 0.1 G from the characteristics of the excitation circuit and coil. The electric field amplitude is then $E \sim 0.02$ V/cm.

We performed complementary microwave (linear) conductivity experiments using a resonant cavity perturbation technique that was previously described[53] and extensively used to study cuprate superconductors[8] (see also Supplementary Notes). The sample was mounted in an evacuated elliptical microwave cavity made of copper and immersed in a liquid helium bath. The complex conductivity of the sample was obtained by measuring the temperature dependence of the $Q$-factor and resonant frequency of the cavity using a microwave demodulator. Similarly to the nonlinear conductivity experiment, the cavity was kept at constant temperature, while the sample temperature was varied in a wide range. We obtained the superconducting response above $T_c$ by subtracting the conductivity measured with an external magnetic field of 16 T (perpendicular to the CuO$_2$ planes) from the zero-field conductivity. No appreciable difference in conductivity was observed between 12 T and 16 T in the relevant temperature range.

## Acknowledgments

We thank A. V. Chubukov for comments on the manuscript. D. P., M. V., M. S. G. and M. P. acknowledge funding by the Croatian Science Foundation under grant no. IP-11-2013-2729. The work at the University of Minnesota was funded by the Department of Energy through the University of Minnesota Center for Quantum Materials under DE-SC-0016371. The work at the TU Wien was supported by FWF project P27980-N36 and the European Research Council (ERC Consolidator Grant No 725521). We acknowledge M. K. Chan for contributing to Hg1201 sample preparation and characterization.

## Author Contributions

DP and MV built the nonlinear conductivity setup, performed measurements, and analyzed the data. MSG and MP built the microwave conductivity setup and performed linear conductivity measurements. MP supervised all conductivity experiments. MP and NB initiated the paraconductivity studies. DP, GY, MG and NB conceived the idea to pursue the percolation-based data analysis. DP performed the percolation calculations. TS prepared the LSCO samples. DP prepared the YBCO-Zn sample. GY and NB prepared the Hg1201 sample. DP, MG and NB wrote the paper with input from all authors.

## References

1.  Keimer, B., Kivelson, S. A., Norman, M. R., Uchida, S. & Zaanen, J., From quantum matter to high-temperature superconductivity in copper oxides. *Nature* **518,** 179-186 (2015).

2.  Corson, R., Mallozzi, L., Orenstein, J., Eckstein, J. N. & Božović, I., Vanishing of phase coherence in underdoped Bi$_2$Sr$_2$CaCu$_2$O$_{8+\delta}$. *Nature* **398,** 221-223 (1999).

3.  Bilbro, L. S. et al., Temporal correlations of superconductivity above the transition temperature in La$_{2-x}$Sr$_x$CuO$_4$ probed by terahertz spectroscopy. *Nature Phys.* **7,** 298-302 (2011).

4.  Rullier-Albenque, F., Alloul, H. & Rikken, G., High-field studies of superconducting fluctuations in high-$T_c$ cuprates: evidence for a small gap distinct from the large pseudogap. *Phys. Rev. B* **84,** 014522 (2011).

5.   Chang, J. et al., Decrease of upper critical field with underdoping in cuprate superconductors. *Nature Phys.* **8,** 751-756 (2012).

6.   Xu, Z. A., Ong, N. P., Wang, Y., Kakeshita, T. & Uchida, S., Vortex-like excitations and the onset of superconducting phase fluctuation in underdoped La$_{2-x}$Sr$_x$CuO$_4$. *Nature* **406,** 486-488 (2000).

7.   Wang, Y., Li L. & Ong, N. P., Nernst effect in high-Tc superconductors. *Phys. Rev. B* **73,** 024510 (2005).

8.   Grbić, M. S. et al., Temperature range of superconducting fluctuations above $T_c$ in YBa$_2$Cu$_3$O$_{7-\delta}$ single crystals. *Phys. Rev. B* **83,** 144508 (2011).

9.   Wang, Y. et al., Field-enhanced diamagnetism in the pseudogap state of the cuprate Bi$_2$Sr$_2$CaCu$_2$O$_{8+\delta}$ superconductor in an intense magnetic field. *Phys. Rev. Lett.* **95,** 247002 (2005).

10.   Yu, G. et al., Universal superconducting fluctuations and the implications for the phase diagram of the cuprates. http://arxiv.org/abs/1210.6942 (2012).

11.   Dubroka, A. et al., Evidence of a precursor superconducting phase at temperatures as high as 180 K in RBa$_2$Cu$_3$O$_{7-\delta}$ (R = Y, Gd, Eu) superconducting crystals from infrared spectroscopy. *Phys. Rev. Lett.* **106,** 047006 (2011).

12.   Boyer, M. C. et al., Imaging the two gaps of the high-temperature superconductor Bi$_2$Sr$_2$CuO$_{6+x}$. *Nature Phys.* **3,** 802-806 (2007).

13.   Sonier, J. E. et al., Inhomogeneous magnetic-field response of YBa$_2$Cu$_3$O$_y$ and La$_{2-x}$Sr$_x$CuO$_4$ persisting above the bulk superconducting transition temperature. *Phys. Rev. Lett.* **101,** 117001 (2008).

14.   Šopík, B., Chaloupka, J., Dubroka, A., Bernhard, C. & Munzar, D., Evidence for precursor superconducting pairing above $T_c$ in underdoped cuprates from an analysis of the in-plane infrared response. *New J. Phys.* **17,** 053022 (2015).

15.   Anderson, P. M., The resonating valence bond state in La$_2$CuO$_4$ and superconductivity. *Science* **235,** 1196 (1987).

16.   Lee, P.A., Nagaosa, N. & Wen, X.G., Doping a Mott insulator: physics of high-temperature superconductivity. *Rev. Mod. Phys.* **78,** 17 (2006).

17. Emery, V. J. & Kivelson, S. A., Importance of phase fluctuations in superconductors with small superfluid density. *Nature* **374,** 434 (1995)**.**

18. Lee, P. A., Amperean pairing and the pseudogap phase of cuprate superconductors. *Phys. Rev. X* **4,** 031017 (2014).

19. Eisaki, H. et al., Effect of chemical inhomogeneity in bismuth-based copper oxide superconductors. *Phys. Rev. B* **69,** 064512 (2004).

20. Stauffer, D. & Aharony, A., *Introduction to Percolation Theory.* (Taylor & Francis, London, 1994).

21. Sahimi, M., *Applications of Percolation Theory.* (Taylor & Francis, London, 1994).

22. Reber, T. J. et al., Prepairing and the filling gap in the cuprates from the tomographic density of states. *Phys. Rev. B* **87,** 060506(R) (2013).

23. Došlić, M., Pelc, D. & Požek, M.*,* Contactless measurement of nonlinear conductivity in the radio-frequency range. *Rev. Sci. Instrum.* **85,** 073905 (2014).

24. Hou, S. L. & Bloembergen, N., Paramagnetoelectric effects in $NiSO_4 \cdot 6H2O$, *Phys. Rev.* **138,** A1218 (1965).

25. Barišić, N. et al., Demonstrating the model nature of the high temperature superconductor $HgBa_2CuO_{4+\delta}$. *Phys. Rev. B* **78,** 054518 (2008).

26. Chan, M. K. et al., In-plane magnetoresistance obeys Kohler's rule in the pseudogap phase of cuprate superconductors. *Phys. Rev. Lett.* **113,** 177005 (2014).

27. Pelc, D., Požek, M., Despoja, V. & Sunko, D., Mechanism of metallization and superconductivity suppression in $YBa_2(Cu_{0.97}Zn_{0.03})_3O_{6.92}$ revealed by [67]Zn NQR. *New J. Phys.* **17,** 083033 (2015).

28. Dorsey, A. T., Linear and nonlinear conductivity of a superconductor near $T_c$. *Phys. Rev. B* **43,** 7575 (1991).

29. Singer, P. W., Hunt, A. W. & Imai, T., [63]Cu NQR evidence for spatial variation of hole concentration in $La_{2-x}Sr_xCuO_4$. *Phys. Rev. Lett.* **88,** 047602 (2002).

30. Bobroff, J. et al., Absence of static phase separation in the high-$T_c$ cuprate $YBa_2Cu_3O_{6+y}$. *Phys. Rev. Lett.* **89,** 157002 (2002).

31. Rybicki, D. et al., $^{63}$Cu and $^{199}$Hg NMR study of HgBa$_2$CuO$_{4+\delta}$ single crystals. *arxiv*:1208.4690 (2012).

32. Alldredge, J. W., Fujita, K., Eisaki, H., Uchida, S. & McElroy, K., Universal disorder in Bi$_2$Sr$_2$CaCu$_2$O$_{8+x}$. *Phys. Rev. B* **87,** 104520 (2013).

33. Kresin, V. Z., Ovchinnikov, Y. N. & Wolf, S. A., Inhomogeneous superconductivity and the pseudogap state of novel superconductors. *Phys. Rep.* **431,** 231-259 (2006).

34. Muniz, R. A. & Martin, I., Method for detecting superconducting stripes in high-temperature superconductors based on nonlinear resistivity measurements. *Phys. Rev. Lett.* **107,** 127001 (2011).

35. Sonier, J. E. et al., Hole-doping dependence of the magnetic penetration depth and vortex core size in YBa$_2$Cu$_3$O$_y$: Evidence for stripe correlations near 1/8 hole doping. *Phys. Rev. B* **76,** 134518 (2007).

36. Grissonnanche, G. et al., Direct measurement of the upper critical field in cuprate superconductors. *Nature Comm.* **5,** 3280 (2014).

37. Perfetti, L. et al., Ultrafast dynamics of fluctuations in high-temperature superconductors far from equilibrium. *Phys. Rev. Lett.* **114**, 067003 (2015).

38. Tinkham, M., *Introduction to Superconductivity.* (McGraw-Hill, New York, 1996).

39. Jaeger, H. M., Haviland, D. B., Orr, B. G. & Goldman, A. M., Onset of superconductivity in ultrathin granular metal films. *Phys. Rev. B* **40,** 182-196 (1989).

40. Yazdani, A. & Kapitulnik, A., Superconducting-insulating transition in two-dimensional α-MoGe thin films. *Phys. Rev. Lett.* **74**, 3037 (1995).

41. Cabo, L., Soto, F., Ruibal, M., Mosqueira, J. & Vidal, F., Anomalous precursor diamagnetism at low reduced magnetic fields and the role of $T_c$ inhomogeneities in the superconductors Pb$_{55}$In$_{45}$ and underdoped La$_{1.9}$Sr$_{0.1}$CuO$_4$. *Phys. Rev. B* **73,** 184529 (2006).

42. Orenstein, J. Optical conductivity and spatial inhomogeneity in cuprate superconductors, in *Handbook of high-Tc superconductivity* (Springer, New York, 2007)

43. Phillips, J. C., Saxena, A. & Bishop, A. R., Pseudogaps, dopants, and strong disorder in cuprate high-temperature superconductors. *Rep. Prog. Phys.* **66,** 2111-2182 (2003).

44.  Barišić, N. et al., Universal quantum oscillations in the underdoped cuprate superconductors. *Nat. Phys.* **9**, 761-764 (2013).

45.  Bangura, A. F. et al., Fermi surface and electronic homogeneity of the overdoped cuprate superconductor $Tl_2Ba_2CuO_{6+\delta}$ as revealed by quantum oscillations. *Phys Rev. B* **82**, 140501(R) (2010).

46.  Poccia, N. et al., Percolative superconductivity in $La_2CuO_{4.06}$ by lattice granularity patterns with scanning micro x-ray absorption near edge structure. *Appl. Phys. Lett.* **104**, 221903 (2014).

47.  Honma, T. & Hor, P. H. Quantitative connection between the nanoscale electronic inhomogeneity and the pseudogap of $Bi_2Sr_2CaCu_2O_{8+\delta}$ superconductors. *Physica C* **509**, 11-15 (2015).

48.  Pelc, D., Popčević, P., Yu, Požek, M., G., Greven, M. & Barišić, N., Unusual behaviour of cuprates explained by heterogeneous charge localization. *Preprint* (2017).

49.  Li, Q., Hücker, M., Gu, G. D., Tsvelik, A. M. & Tranquada, J. M., Two-dimensional superconducting fluctuations in stripe-ordered $La_{1.875}Ba_{0.125}CuO_4$. *Phys. Rev. Lett.* **99**, 067001 (2007).

50.  Sasagawa, T., Kishio, K., Togawa, Y.,  Shimoyama, J. & Kitazawa, K. First-order vortex-lattice phase transition in $(La_{1-x}Sr_x)_2CuO_4$ single crystals: universal scaling of the transition lines in high-temperature superconductors. *Phys. Rev. Lett.* **80**, 4297 (1998).

51.  Barišić, N. et al., Hidden Fermi-liquid behavior throughout the phase diagram of the cuprates. http://arxiv.org/abs/1507.07885 (2015).

52.  Pelc, D., Vučković, M., Grafe, H.-J., Baek, S.-H. & Požek, M., Unconventional charge order in a co-doped high-$T_c$ superconductor. *Nature Comm.* **7**, 12775 (2016).

53.  Nebendahl, B., Peligrad, D.-N., Požek, M., Dulčić, A. & Mehring, M., An ac method for the precise measurement of Q-factor and resonance frequency of a microwave cavity. *Rev. Sci. Instrum.* **72**, 1876 (2001).

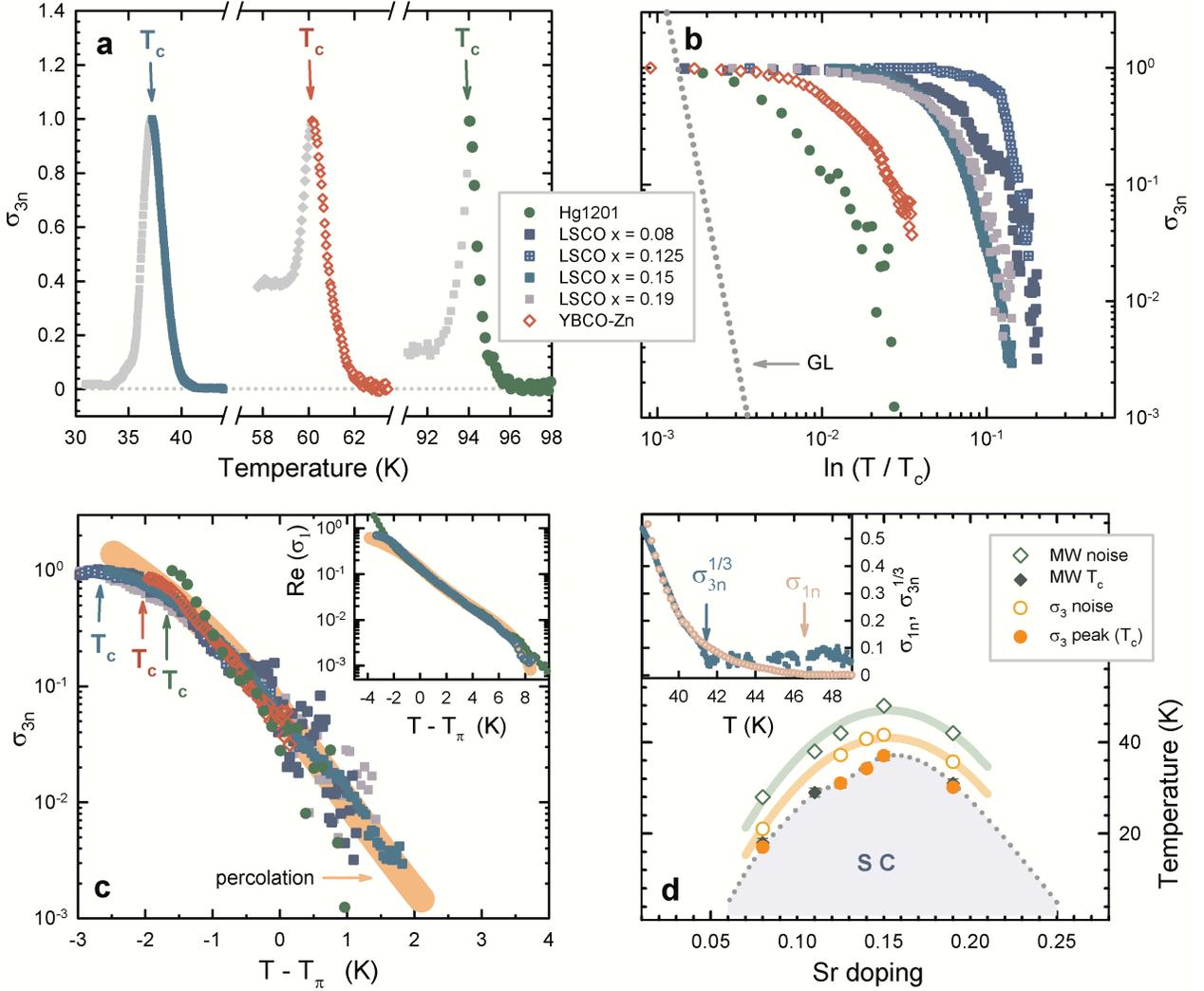

**Fig. 1 | Temperature dependence of in-plane linear and nonlinear response in the cuprates.**
**a,** Nonlinear conductivity $\sigma_{3n}$ ($\sigma_3$ normalized to its peak value, which corresponds to the bulk $T_c$ indicated by arrows) for three representative samples close to optimal hole doping: Hg1201, YBCO-Zn, and LSCO-0.15 ($x = 0.15$). **b,** Nonlinear conductivity vs. Ginzburg-Landau reduced temperature $\ln(T/T_c)$, demonstrating that $T_c$ is not the common scale for superconductivity emergence. Dotted line is the GL prediction. **c,** $\sigma_{3n}$ shifted by a sample-dependent temperature $T_\pi$ collapses to a single curve. This demonstrates the existence of a universal emergence temperature/energy scale. $T_c$ is indicated by arrows (LSCO-0.15, LSCO-0.19 and LSCO-0.08 samples have indistinguishable $T_c$ values on this scale; $T_c$ for LSCO-0.125 is at $T - T_\pi = -4.5$ K). Orange line is the prediction of the simple site-percolation model discussed in the text. Inset:

linear conductivity of Hg1201 and LSCO-0.15 along with the model prediction. **d,** Phase diagram of LSCO with focus on the characteristic temperatures below which the superconducting response is first resolved in both linear and nonlinear conductivity - errors are determined from the root-mean-square noise level and are within the symbol size. These temperatures are significantly different despite the similar signal-to-noise ratio of $\sim 10^3$ (at $T_c$) of the two experiments, in agreement with the percolation model. The positions of the peaks in $\sigma_{3n}$ and the real part of $\sigma_1$ give consistent $T_c$ values. Error bars for $T_c$ are given by the peak width (and are within the symbol size for $\sigma_{3n}$). Lines are guides to the eye. The inset shows a comparison of $\sigma_{1n}$ and $\sigma_{3n}^{1/3}$ for LSCO-0.15, demonstrating that $\sigma_{3n} \propto \sigma_{1n}^3$ until the third order response is indistinguishable from noise. The two characteristic temperatures are marked with arrows.

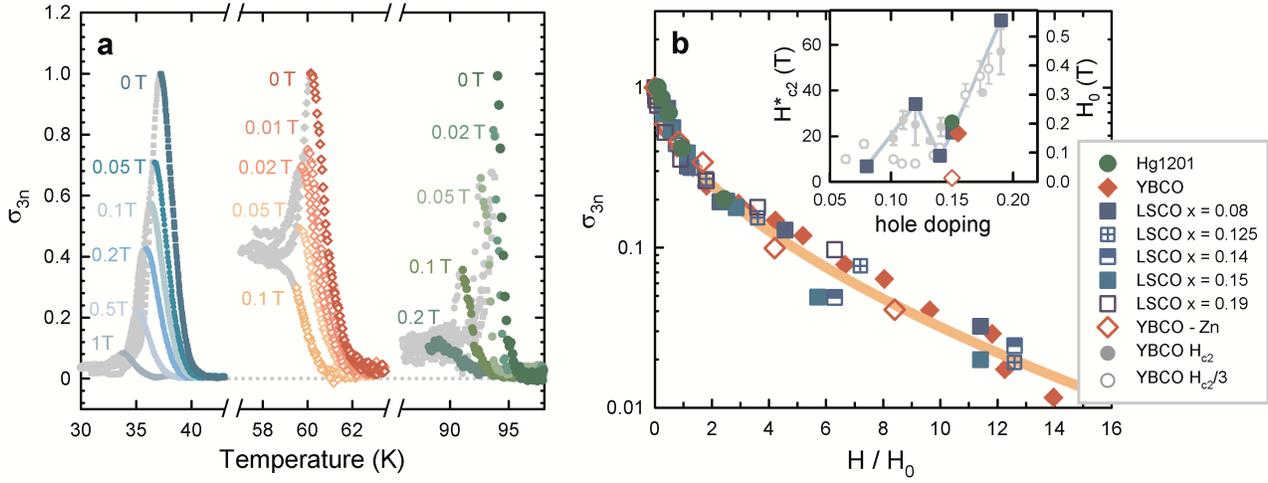

**Fig. 2 | Influence of external c-axis magnetic field on in-plane nonlinear response. a,** Nonlinear conductivity at various applied fields for Hg1201, YBCO-Zn and LSCO-0.15, normalized to the zero-field peak values. **b,** Scaling of the nonlinear conductivity with magnetic field. Symbols indicate peak values of $\sigma_3$ upon subtracting the high-field step and normalizing to the zero-field values. The orange line is the result of an effective medium calculation for the percolation model (see text). Inset: suppression field $H_0$ compared with previous results for $H_{c2}$ for YBCO obtained using muon spin rotation[35] (full circles) and transport[36] (empty circles). The grey line is a guide to the eye. Errors (1 s.d.) in $\sigma_{3n}$ and $H_0$ are within the symbol size.

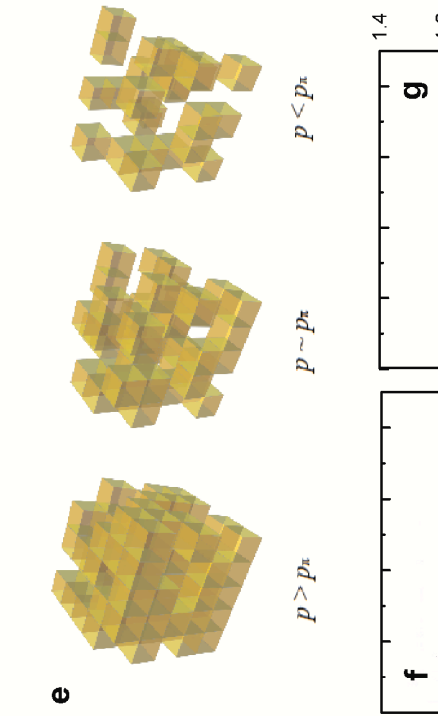
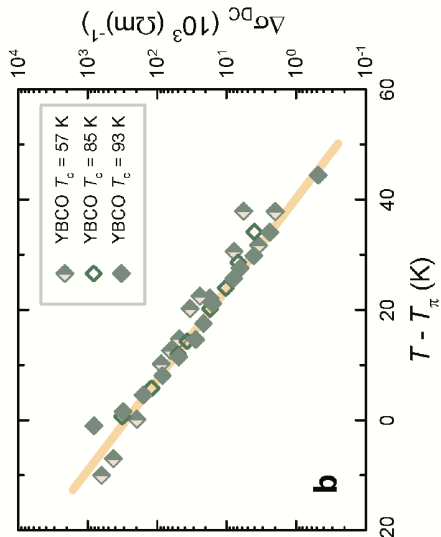
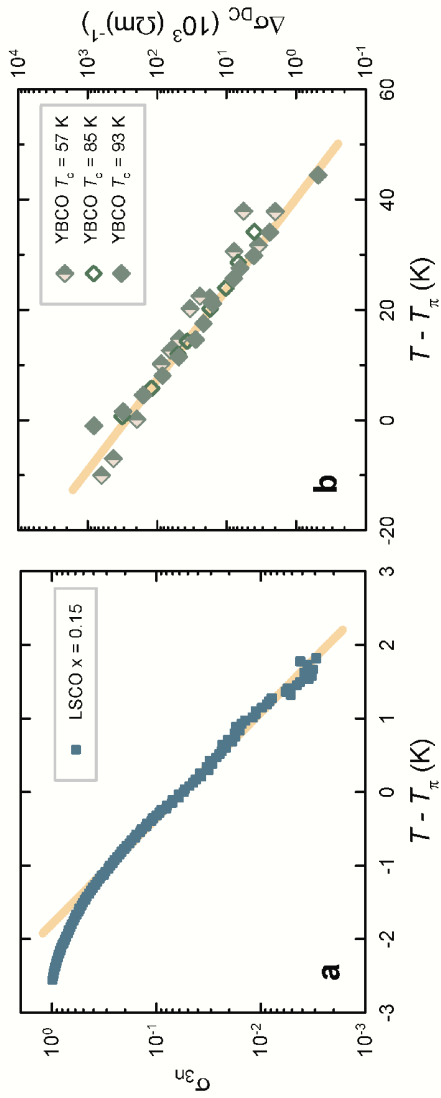
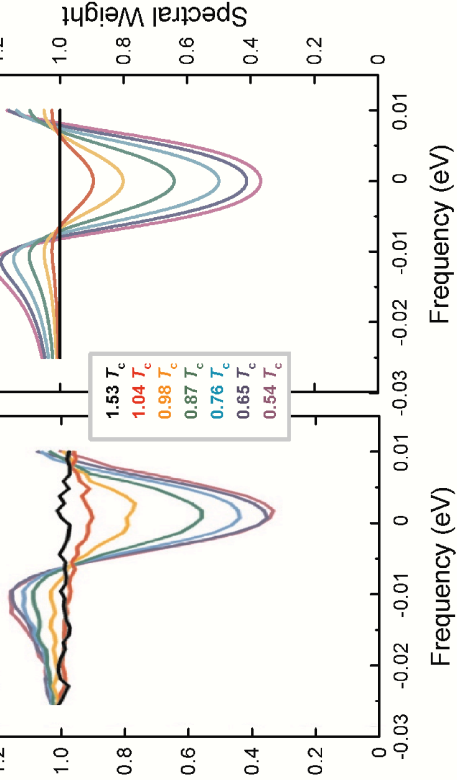
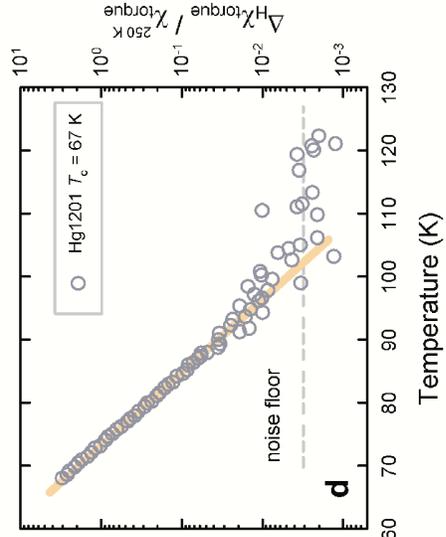
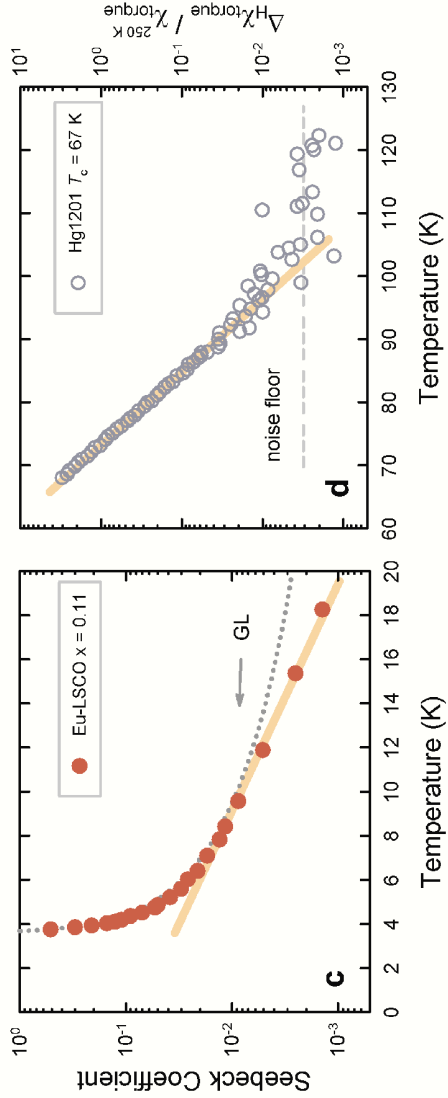

**Fig. 3 | Universal percolation physics in the cuprates. a-d,** Four different experimental probes show qualitatively similar non-GL behaviour (full lines) above $T_c$: **a,** nonlinear conductivity of optimally-doped LSCO (this work); **b,** superconducting contribution to DC conductivity of YBCO, adapted from ref. 4; **c,** Seebeck coefficient of Europium co-doped LSCO ($x = 0.11$) adapted from ref. 5 - the dotted line is the 2D GL prediction valid within 2 K of $T_c$; **d,** torque magnetometry in underdoped Hg1201 ($T_c = 67$ K) - the dashed line indicates the noise floor. Adapted from ref. 10. The decay constants of different observables are proportional to the universal scale $T_0$, but with different prefactors that can be directly calculated from the suggested model (see Supplementary Notes). **e,** Schematic representation of the superconducting site percolation model in a cubic geometry. Yellow patches are superconducting. Above the critical concentration $p_\pi$ (below $T_\pi$), a sample-spanning superconducting cluster exists. **f,** Measured tomographic density of states for optimally doped $Bi_2Sr_2CaCu_2O_{8+x}$ adapted from ref. 22. **g,** Calculated density of states, assuming a Gaussian gap distribution with full width of 3.2 meV (see Supplementary Notes). In **a-d**, the lines are not the result of a direct calculation, but rather highlight pure exponential decay.

| Sample | $p$ | $T_c$ | $T_0$ |
|--------|-----|-------|-------|
| Hg1201 | 0.14 | 94.0 K | 25.1 ± 1.4 K |
| LSCO-0.08 | 0.08 | 17.3 K | 29.9 ± 1.4 K |
| LSCO-0.125 | 0.125 | 31.0 K | 28.2 ± 0.9 K |
| LSCO-0.15 | 0.15 | 37.2 K | 28.6 ± 0.2 K |
| LSCO-0.19 | 0.19 | 30.1 K | 29.3 ± 1.5 K |
| YBCO-Zn | 0.15 | 60.2 K | 26.1 ± 0.6 K |

**Table 1 | Hole doping level $p$, superconducting transition temperature $T_c$, and percolation scale $T_0$ for the investigated samples.** For LSCO, $p$ = x, whereas for Hg1201 and YBCO-Zn, the estimate is based on the findings in refs. 25 and 27, respectively. As described in the text, $T_c$ corresponds to the peak in the nonlinear conductivity; we estimate the error to be less than 1%. $T_0$ is obtained from nonlinear conductivity using a 3D site percolation model.

Supplementary Materials for

**"Emergence of superconductivity in the cuprates via a universal percolation process"**

Damjan Pelc,[1,†,#] Marija Vučković,[1,†] Mihael S. Grbić,[1] Miroslav Požek,[1,*] Guichuan Yu,[2] Takao Sasagawa,[3] Martin Greven,[2,*] Neven Barišić[2,4,*]

[1]Department of Physics, Faculty of Science, University of Zagreb, Bijenička 32, HR-10000, Zagreb, Croatia

[2]School of Physics and Astronomy, University of Minnesota, Minneapolis, MN 55455, USA

[3]Materials and Structures Laboratory, Tokyo Institute of Technology, Kanagawa 226-8503, Japan

[4]Institute of Solid State Physics, TU Wien, 1040 Vienna, Austria

†These authors have contributed equally

#Present address: [2]School of Physics and Astronomy, University of Minnesota, Minneapolis, MN 55455, USA

*Correspondence to: mpozek@phy.hr, greven@umn.edu, neven.barisic@tuwien.ac.at

**Content**

Supplementary Methods

Supplementary Notes

Figures S1-S6

Supplementary references

## Supplementary Methods

Nonlinear response measurements typically require relatively large applied fields in order to detect the small signals. Hence the most serious problem plaguing nonlinear measurements of conductive systems is Joule heating, i.e., the variation of the conductivity/susceptibility with temperature induced by resistive heating of the sample. If a constant or slowly varying electric field is used to detect nonlinear response, the large current will heat the sample during measurement, and spurious nonlinear contributions will appear if the resistance depends on temperature. Conventionally, millisecond field pulses are used to alleviate the heating problem, but heating still plays a role for highly conducting samples and needs to be disentangled from other possible contributions[1,2]. A pivotal step in our experiment is the use of a high-frequency excitation field – if the frequency is high enough (17 MHz in our case), the time-dependent temperature change of bulk samples cannot follow the rapidly changing field, and no heating-induced nonlinear signal is observed. Of course, an average, time-independent heating is still present, but this does not influence the measurement of the nonlinear response. In principle, time-independent heating may cause a small shift of the sample temperature, but this was determined to be negligible in our case from a comparison of $T_c$ (peak positions) in linear and nonlinear conductivity throughout the phase diagram of LSCO. The intrinsic nonlinear response is measured at a harmonic of the fundamental excitation frequency, which enabled high sensitivity due to phase sensitive detection

An essential feature of our experiment is that it is performed without contacts on the sample, thereby eliminating possible nonlinearities in contact regions. Moreover, the contact-free measurement enables control of the sample temperature in a wide range without changing the temperatures of the excitation and detection coils. A special sample probe for use in superconducting magnets was constructed for this purpose. We note that prior nonlinear microwave experiments[3-5], performed on YBCO, were interpreted either phenomenologically or their importance for studying the emergence regime was not recognized. Furthermore, our MHz-range experiment is much closer to the zero-frequency limit than these prior studies, which minimizes the need for possible corrections from frequency-dependent effects.

**Supplementary Notes**

## 1. In-depth comparison of nonlinear response to vortex and GL theory

Several mechanisms are known to give rise to nonlinear response in the vicinity of $T_c$: vortex-related effects, superconducting fluctuations, and inhomogeneity/percolation physics. Here we attempt a detailed comparison of the temperature-dependent third-harmonic signal to predictions for these mechanisms in order to elucidate the origin of the nonlinearity.

**Vortex dynamics.** The first potential explanation – vortex dynamics – has been extensively studied in type-II superconductors due to its technical importance. Linear and nonlinear harmonic responses have been calculated using phenomenological vortex electrodynamics appropriate for cuprate superconductors[6,7]. However, our results are incompatible with the calculations, for several reasons. According to those calculations, amplitudes of higher harmonics should diminish rapidly – third harmonic response should be significantly smaller than second harmonic (in a small constant external field of about $H_{c1} \sim 10$ G). In order to test this, we measured the second harmonic response, applying both an oscillating and a constant magnetic field (in the range from about 1 to 100 G). We only find a detectable second-harmonic peak in the cleanest sample, Hg1201, its width an order of magnitude smaller than the third harmonic, and the signal already below the detection limit in an external field of about 100 Gauss. This weak second-harmonic peak (orders of magnitude smaller than the third-harmonic peak) just above $T_c$ is only detected in Hg1201, probably because of the fact that this cuprate exhibits relatively weak vortex pinning[8]; LSCO and YBCO-Zn feature prominent point-like disorder that can pin vortices and diminish their contribution to the nonlinear response. Moreover, all harmonics are predicted to diminish equally quickly with applied external field, and to disappear above the superconducting $T_c$. This prediction indeed describes a separate reference measurement that we performed on a niobium single crystal (Fig. S1), but it is inconsistent with the observed universal third-harmonic behaviour in the cuprates. Although it is possible that a comparatively small vortex contribution is present very close to $T_c$ (especially in Hg1201), where our minimal percolation model is invalid, the significant signal observed to relatively high temperatures cannot be due to vortices.

**Ginzburg-Landau theory.** The second possible origin of the nonlinearity above $T_c$ are classic Ginzburg-Landau (GL) superconducting fluctuations, which have been extensively investigated in the cuprates using linear response in a wide frequency range[9-12]. Nonlinear response is a better probe of fluctuation contributions, since in linear response one must always attempt to determine and subtract a normal-state contribution, a complication that is absent in the third harmonic, as demonstrated in the main text. At a quantitative level, the linear and nonlinear GL-fluctuation response has been calculated beyond mean field[13] and with included anisotropy[9,12] (only linear conductivity). For an isotropic type-II superconductor, the nonlinear conductivity in both the Gaussian and critical fluctuation regimes is shown to be proportional to the linear conductivity, as follows. In general, one can define a field-dependent conductivity, $\sigma(E) = \sigma(E = 0)\Sigma(E)$. The scaling function $\Sigma(E)$ has different forms in different fluctuation regimes and for small/large electric fields, but in the small-field approximation the leading term is always $1 + A(E/E_0)^2$, where $A$ is a numerical constant and $E_0$ a reference electric field[13]. The field $E_0$ depends on temperature through the mean-field correlation length $\xi(T)$, as $E_0 \propto \xi^{-3}$. Therefore, $\sigma_3 = A/E_0^2 \sigma_1 \propto \sigma_1 \xi^6$. Since, in GL theory, the linear and nonlinear responses are due to the same fluctuation physics, such a scaling relationship between the two should hold regardless whether or not $ab$-plane/$c$-axis anisotropy and short-wavelength cutoffs[9,12] are included. Thus one can directly compare the temperature dependence of the linear/nonlinear response to the predictions of linear GL fluctuation theory. Due to the low frequency of our experiment, the linear response simply corresponds to the in-plane dc linear conductivity, $\sigma_{ab}^{DC} = \sigma_1 \propto \sigma_3/\xi_{ab}^6$. The dc conductivity is given by[9]

$$\sigma_{ab}^{DC} = \frac{e^2}{16\pi\hbar\xi_{0c}}\left(\frac{\xi_{ab}(T)}{\xi_{0ab}}\right)^{z-1} f(Q_{ab}, Q_c) \qquad (1)$$

where $\xi$ is the superconducting coherence length, the indices $ab$ and $c$ correspond to in-plane and $c$-axis quantities, respectively, $z$ is the dynamical exponent, and $f(Q_{ab}, Q_c)$ a function of the temperature-dependent anisotropic fluctuation cutoffs $Q_{ab}$ and $Q_c$ in reciprocal space. The cutoffs are $Q_{ab,c} = \sqrt{3}\Lambda_{ab,c}\xi_{ab,c}(T)/\xi_{0ab,c}$, where $\Lambda_{ab,c}$ are temperature-independent cutoff scales[9]. Since the electric fields applied in our measurements are small, they are significantly below $E_0$ (except perhaps in the closest vicinity of $T_c$, where $E_0$ rapidly goes to zero).

The relevant dimensionless temperature variable for $\xi(T)$ in GL theory is $\ln(T/T_c)$, and data for several cuprates are plotted versus this GL reduced temperature in Figs. 1b and S2. The theoretical prediction obtained from Eq. (1) using the realistic parameters[11] $\Lambda_{ab} = 0.1$ and $\Lambda_c = 0.02$ is shown in Fig. 1b. The theoretical prediction clearly decays much faster than the data for all investigated samples. We note that the choice of a different value for $T_c$ cannot improve the agreement between data and theory. This is demonstrated in Fig. S2 for the case of LSCO with $x = 0.15$ (measured $T_c = 37.2$ K). Better agreement can be obtained if the reduced temperature variable is multiplied by material-dependent constants for different samples, but in the case of GL fluctuations these would be additional arbitrary non-universal free parameters without obvious physical meaning. Even more importantly, the shape of the temperature dependence cannot be satisfactorily reproduced by GL theory, whatever scaling one employs on the temperature axis.

**Relevance of 2D fluctuations.** We note that the quasi-2D case of Eq. (1) – corresponding to Kosterlitz-Thouless physics – is readily obtained from first principles[9,12] or by setting[11] $\Lambda_{ab} \gg \Lambda_c$. In that case, the agreement with measurements is even worse than in the anisotropic 3D case, which indicates that the prominent transport nonlinearities observed in $La_{1.875}Ba_{0.125}CuO_4$ (LBCO-1/8) and attributed to quasi-2D Kosterlitz-Thouless physics[14] are exceptional – nonlinear response cannot be said to universally originate from 2D fluctuations in the cuprates. A similar conclusion can be reached upon reexamining recent Seebeck coefficient data[15] for underdoped $La_{2-x-y}Eu_ySr_xCuO_4$ (Eu-LSCO). Although in the vicinity of $T_c$ the emergence of superconductivity seems to be well described by a 2D Gaussian model[15], when the very same data are plotted on a linear temperature scale (Fig. 3c in the main text) a clear exponential tail is observed that extends to higher temperatures. Importantly, in Eu-LSCO the superconducting transition is strongly suppressed and broadened[15] compared to LSCO, with $T_c$ on the order of 5 K, rendering Eu-LSCO a rather unrepresentative cuprate. Its low $T_c$ provides for a large accessible fluctuation range in *relative* temperature $(T - T_c)/T_c$, but on the absolute temperature scale the fluctuations extend only about $T_0/2 \sim 15$ K above $T_c$, as in all other cuprates examined in our work. The experiments on Eu-LSCO are thus compatible with our conclusion that conventional GL fluctuations (with $T_c$ as the relevant scale) are only visible very close to $T_c$, whereas further away from $T_c$ inhomogeneity-induced superconducting percolation (with the universal $T_0$ as the relevant scale) dominates. This is also clear from Fig. 1c in the main text:

close to $T_c$ the percolation prediction deviates from the data, as expected when the fraction of superconducting patches is large. Yet in that regime the 2D theory as applied to Eu-LSCO may not be universally valid in the cuprates: Eu-LSCO has many similarities to LBCO-1/8, where charge stripes lead to dynamical layer decoupling and effective two-dimensionality[14,15]. Similar effects could plausibly occur in Eu-LSCO, and while certainly interesting in their own right, they cannot be said to be representative of all cuprates.

## 2. Details of the minimal percolation model

The main idea of the model is that nanoscale superconducting patches form and proliferate in the material (Fig. 3e), with macroscopic superconductivity then emerging via a percolation process. We assume perfectly connected square or cubic patches (two-dimensional (2D) or three-dimensional (3D) nearest-neighbor site percolation) that are either non-superconducting, each with a normal resistance $R_n$, or superconducting, each with a nonlinear resistance[16]

$$R_s\left(j\right) = R_0 + \frac{R_n - R_0}{e^{-4\left(j - J_c\right)} + 1} \qquad (2)$$

where $j$ is the current through the superconducting patch, $R_0$ its residual resistivity (due to the finite size of the patch, and $R_0 << R_n$), and $J_c$ the patch critical current; $j$ and $J_c$ are dimensionless currents. We assume the patches to be static (which is probably not a good approximation well above $T_c$) and neglect Josephson couplings (not a good approximation very close to $T_c$). The fraction of superconducting patches is taken to be $P$, with $P \rightarrow 0$ at high temperatures and $P \rightarrow 1$ well below $T_c$. The critical concentration at which the system percolates, $P_\pi$, depends on the dimensionality of the system and the chosen percolation model. In the nearest-neighbor site percolation scenario we use here, we have[17] $P_\pi \approx 0.3$ (3D) and $P_\pi \approx 0.6$ (2D). Site percolation is physically realistic in the case of superconducting patches with different local $T_c$ values, as seen in STM[18-20], but the particular choice of the percolation model does not critically affect the modeling, as we show below. In order to make a quantitative comparison to experiment, a dependence of $P$ on temperature must be assumed. The simplest possibility is a linear dependence, $P_\pi - P = (T - T_\pi)/T_0$, where $T_0$ is the universal temperature scale that connects $P$ and $T$. Physically, the linear dependence is equivalent to taking the distribution of local superconducting gaps to be a simple boxcar function of width $T_0$. Yet the linear term is the leading term for any realistic distribution, and thus this approximation is always valid not too far from the $T_c$. Our goal, in the spirit of the minimal model, is to avoid any assumptions related to the gap distribution. This approach in case of linear and nonlinear conductivity gives good results. We illustrate the difference between our assumption of a linear dependence of $P$ on $T$ and a more realistic Gaussian distribution of local gaps in Fig. S3. At high temperatures, the Gaussian distribution gives better asymptotic behaviour, eliminating the artificial cutoff present in the linear approximation and causing the exponential tails of conductivity, magnetization, etc. Yet in the temperature range where linear and nonlinear conductivity is measurable, the differences are minimal.

The linear and nonlinear responses are calculated via effective medium theory[21], using the form most appropriate for site percolation[22]. In order to calculate the third-order nonlinear conductivity, the dependence of the voltage on current was determined, and $\sigma_3$ was obtained through an expansion in powers of voltage. Due to the percolative nature of the system, $\sigma_3$ is insensitive to the values of $R_0$ and $J_c$ in the region of interest close to $T_\pi$ (as long as the current $j$ is much smaller than $J_c$). Thus the only parameters entering the calculation of $\sigma_3$ are $T_0$ and the percolation threshold concentration $P_\pi$ of superconducting patches (which depends on the number of spatial dimensions, on site vs. bond percolation, etc.). $R_0$ is used in the linear response calculation, and was determined to be $0.005R_n$, which is realistic for nanoscale patches at a finite excitation frequency[16,23].

In order to obtain $T_0$ and to determine if 3D (with $P_\pi \approx 0.3$) or 2D (with $P_\pi \approx 0.6$) site percolation is more appropriate, we simultaneously calculate the linear and nonlinear conductivity and compare to the measurements (Fig. 1c in the main text). Although the results do not critically depend on $P_\pi$, a 3D site percolation model with $P_\pi = 0.31$ yields the best agreement with the data. In particular, it enables the linear and nonlinear response in LSCO to be described with a single $T_0 = 28.0 \pm 0.4$ K, whereas in the 2D model the discrepancy between $T_0$ obtained from linear and nonlinear conductivities differs at least by 25% (Fig. S4). With $P_\pi = 0.31$ fixed, individual fits to only the nonlinear response of all investigated compounds (Table 1) gives the overall estimate $T_0 = 27 \pm 2$ K, whereas the simultaneous calculation of both $\sigma_1$ and $\sigma_3$ for LSCO gives the higher precision above. We emphasize that the parameter $R_0$ does not influence the determination of $T_0$: $R_0$ influences the shape of the linear conductivity curve, whereas $T_0$ sets the range of the superconducting contribution. $T_\pi$ is calculated separately in a model-free way to obtain the best data scaling, with typical uncertainties smaller than 0.05 K. The LSCO-0.15 data are taken as a reference since they have the best signal-to-noise ratio. Yet since $T_0$ essentially gives the decay constant of $\sigma_3$ (in our calculation the decay rate is 42.6 K/$T_0$), the determination of $T_0$ is independent of $T_\pi$. The calculated curves depart from measurements close to the macroscopic $T_c$, which is expected – once a significant volume fraction of the sample is superconducting, Josephson couplings can no longer be neglected, macroscopic phase coherence sets in, and the simple percolation picture needs corrections.

One can perform a similar effective medium calculation for 3D nearest-neighbor bond ($P_\pi$ = 0.25) rather than site percolation, or for any other percolation model with a similar critical concentration, and fit to the nonlinear data. This in itself poses no problems and will increase $T_0$ (by about 20%). However, the linear conductivity provides a constraint – similar to the 2D case, it cannot be simultaneously obtained with the same $T_0$ (the difference being about 10%, larger than the uncertainties). Also, the corrections from $P$ vs. $T$ nonlinearity may become important. In any case, the difference between $P_\pi$ = 0.31 and 0.25 is not very significant in view of the crudeness of the modeling, but the data support 3D percolation. The cuprate superconductors are known to be strongly anisotropic; in the site percolation model, this translates to anisotropy within the patches (i.e., they are elongated in the $c$-direction), but this does not change the percolation threshold. Since we measure in-plane response, the threshold is the only important parameter. A possible exception would be systems with effectively decoupled layers (such as Eu-LSCO and LBCO close to doping 1/8, as discussed above).

## 3. Percolation interpretation of the tomographic density of states

Along with transport properties, the percolation model can be used to explain other seemingly unconventional results for the cuprates. One example is the tomographic density of states obtained in recent photoemission measurements[24]. The effective gap obtained in these experiments does not close at the macroscopic $T_c$, but a 'filling' of the density of states is observed to extend to temperatures ~ 1.2 $T_c$ (Fig. 3f in the main text). The gap filling was attributed to an increased superconducting pair-breaking rate, and the response above $T_c$ to pre-formed pairs. However, as we now show, both effects arise naturally if one assumes a spatial gap distribution. In Ref. [24], the density of states was fitted to the standard expression

$$\rho_{Dynes} = \Re \frac{\omega - i\Gamma}{\sqrt{(\omega - i\Gamma)^2 - \Delta^2}} \, , \qquad (3)$$

where $\omega$ is the frequency relative to the Fermi level, $\Gamma$ the pair-breaking rate, and $\Delta$ the superconducting gap. In order to describe the data with this formula, the pair-breaking rate has to increase to $\Gamma \sim \Delta$ close to $T_c$, which signals that the description is no longer physically valid. We find that the experimental result can be quantitatively reproduced by employing a temperature-independent $\Gamma$ and by considering that the experiment measures the average density of states in a system with a real-space gap distribution. We then simply convolute the density of states with a gap distribution function, and employ the standard BCS temperature dependence for the gaps. A Gaussian gap distribution with mean $\Delta_m = 9.6$ meV and full width at half maximum $\Delta_0 = 3.2$ meV (in line with Refs. [18-20] and with our nonlinear conductivity measurements) yields rather good agreement with the experiment at all temperatures (Fig. 3g in the main text). This constitutes a strong, independent confirmation of the percolation/gap disorder scenario.

## 4. Percolation processes in other aspects of cuprate physics

The vast majority of theoretical models applied to the cuprates are based on the assumption of lattice translational invariance. Nevertheless, the effects of various types of disorder have been extensively discussed, including superconducting percolation and pseudogap inhomogeneity[25-37]. Experimental observations indicate that inherent inhomogeneity is prevalent in the cuprates, in related lamellar systems, and in perovskites in general[25-27]. For example, it is known from STM and nuclear quadrupolar resonance measurements[18-20,28-30,36] that significant nanoscale inhomogeneity exists already well above the pseudogap temperature. In fact, it has been argued that transport data are consistent with a temperature-dependent carrier localization that is inhomogeneous in real space[38]. In this picture, the pseudogap temperature $T^*$ signifies a percolation transition involving $CuO_2$ units with one localized hole, and the localization process is complete at the slightly lower temperature $T^{**}$. This is also consistent with an extended analysis of STM data, relating high-energy inhomogeneous gaps to effective pseudogaps and superconductivity[37]. Doping and temperature have similar effects in this picture: the $CuO_2$ sheets evolve from $x$ carriers deep in the pseudogap state to $1 + x$ carriers both at high hole-dopant concentrations and at high temperature. As a function of temperature, the inhomogeneous (de)localization must span a broad range of at least 1000 K in order to be consistent with both Hall effect and STM results[18,36,38]. In stark contrast, the superconducting heterogeneity scale $T_0$ identified in the present work is only about 30 K. Within this picture, the localization of one carrier per $CuO_2$ unit renormalizes the underlying nanoscale inhomogeneity, giving rise to the smaller superconducting inhomogeneity scale $T_0$ (which corresponds to a gap distribution with a width of about 3 meV).

## 5. Nonlinear response in a magnetic field

The field-dependent nonlinear response can be calculated in the percolation model if one assumes a dependence of $J_c$ on external magnetic field. If the simplest, linear dependence is taken, the nonlinear response decreases exponentially with field. Empirically, the form of the $\sigma_{3n}$ $(H/H_0)$ master curve in Fig. 3b is consistent with

$$\sigma_{3n}\left(H/H_0\right) = e^{-\left(H/H_0\right)^{1/2}} \qquad (4)$$

where $H_0$ is the suppression field scale. We note that fitting the curves in in Fig. 3b in the main text with the more general functional form $\sigma_{3n}(H) = \exp\left[-\left(H/H_0\right)^{\beta}\right]$ yields an exponent somewhat larger than 1/2, $\beta = 0.59 \pm 0.07$, and values of $H_0$ shifted upward by about 10% compared to the curve with $\beta = \frac{1}{2}$ (see Fig. S5 for a comparison). Therefore, in order to obtain better agreement with experiment, rather than a linear dependence of the critical current on external field, the phenomenological form $J_c \sim const. - \left(H/H_0\right)^{\beta}$ should be used. However, such a simple treatment ignores the important fact that, due to phase coherence effects, a large superconducting cluster will have a different dependence of $J_c$ on $H$ than a small one. Thus a large field will render most clusters normal, except for the largest ones (including the sample-spanning cluster when $P > P_{\pi}$). This cannot be easily incorporated in the effective medium calculation and would require more elaborate lattice simulations, which we leave for future work. Importantly, due to the percolation physics, cluster-size effects are much more important when considering the magnetic field response than the zero-field gap inhomogeneity close to $T_c$. The cluster sizes vary over a wide range, which implies that cluster size and underlying (average) gaps are only weakly correlated. Importantly, the free energy of a cluster also depends on its size. Therefore large clusters with somewhat smaller average gaps win over small clusters with larger gaps: the relative difference in gaps is roughly $T_0/T_c$, which is, e.g., about 0.3 for optimally doped Hg1201, whereas the cluster sizes can differ by many orders of magnitude.

The experimental determination of nonlinear conductivity peak heights on external field is performed after the subtraction of the field-independent step-like "background" contribution (Fig. S6). For most samples, this step is small compared to the peak, and therefore the correction is minor. However, this is not the case for YBCO-Zn, probably due to the fact that it is a powder sample – in the percolation picture, the signal below $T_c$ in a single crystal is due to the single sample-spanning cluster, whereas in a powder every grain contributes with its own grain-spanning cluster. Thus the relative intensity of the sample/grain-spanning clusters below $T_c$ is significantly larger in a powder. The subtraction is performed by fitting a simple three-parameter sigmoidal form, $\sigma_{3n,step} = a/\left[1 + \exp\left((T - T_{step})/b\right)\right]$, to the high-field data (Fig. S6a) and then subtracting this contribution from the lower-field data. $T_{step}$ is taken to have the same field dependence as $T_c$. We note that the subtraction is not performed in the analysis of the zero-field temperature-dependent data (Fig. 1 in the main text), since the correction is small on a logarithmic scale and only important close to the peak.

A quantitative comparison of our $H_0$ to $H_{c2}$ values from the literature is more problematic than might seem due to the large $H_{c2}$ values typically encountered in the cuprates, and because the complicated temperature-field phase diagram renders an unambiguous identification of $H_{c2}$ difficult. In the inset of Fig. 3b in the main text, we compare our results for $H_0$ with $H_{c2}$ determined from resistivity/heat conductivity measurements[39] and muon spin rotation[40] for YBCO. Without entering into the details of the $H_{c2}$ determination and the assumptions involved there, we conclude that a qualitative agreement between the doping dependencies of $H_0$ and $H_{c2}$ does exist. A more detailed quantitative comparison does not make sense anyway, since our effective medium calculation should only be viewed as a first approximation.

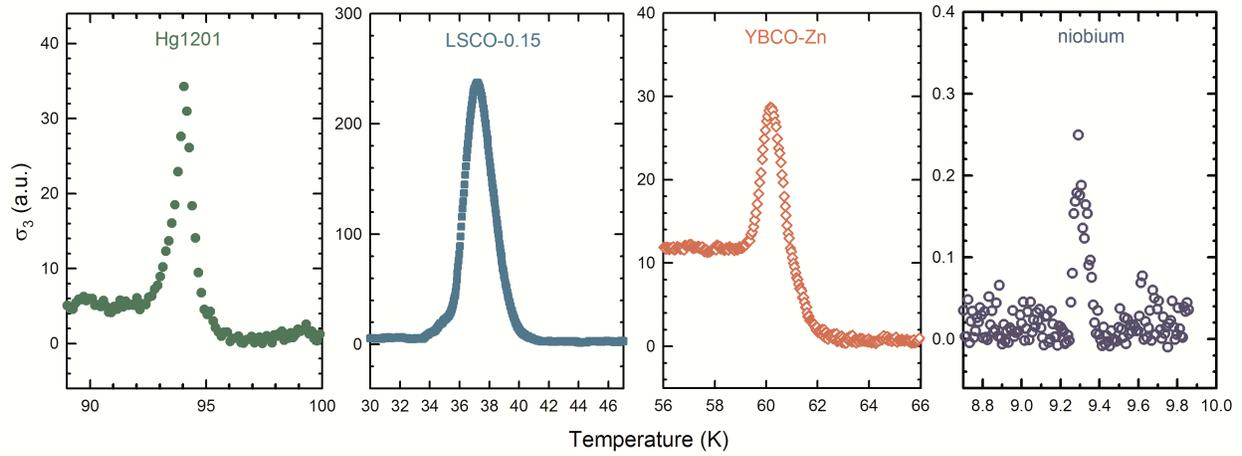

**Fig. S1 |** Raw third-harmonic data for three cuprate samples, compared to a niobium single crystal. Even though a peak is visible for Nb as well, it is substantially narrower and weaker in intensity than for the cuprates – the peak signals for cuprate crystals of similar size are 2-3 orders of magnitude larger. The units are the same for all samples – note the difference in scale for the niobium measurement.

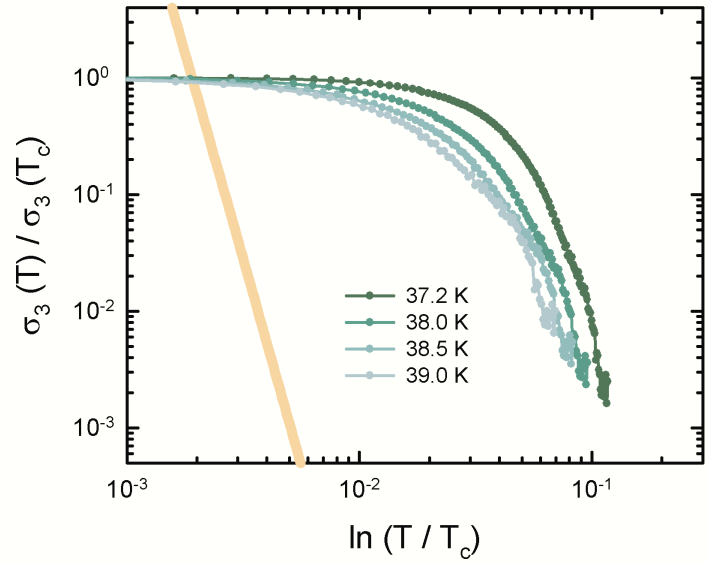

**Fig. S2 |** In-plane nonlinear response of LSCO-0.15, plotted on the Ginzburg-Landau reduced temperature scale, for different choices of $T_c$. The value that corresponds to the peak in $\sigma_{3n}$ is 37.2 K. The yellow line is the theoretical GL prediction, multiplied by a constant factor of 10 in an attempt to improve the agreement.

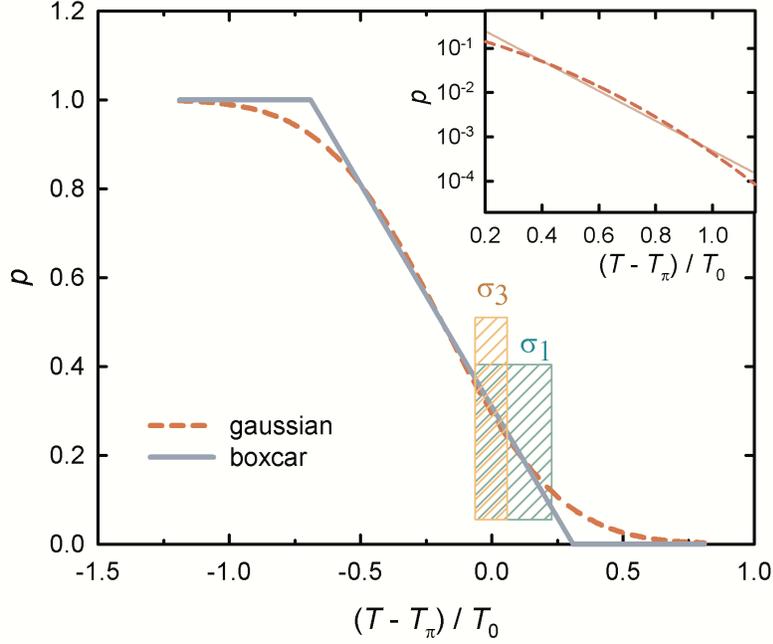

**Fig. S3** | Comparison of two different assumptions for the dependence of the superconducting volume fraction $p$ on temperature, for 3D site percolation. The solid line is the linear assumption - equivalent to a boxcar distribution of local gaps - and the dashed line is obtained if a Gaussian local gap distribution is assumed. The widths of the distributions are the same and equal to $T_0$. The shaded areas indicate the temperature ranges in which linear and nonlinear conductivity are appreciable in the experiment; this demonstrates that the linear assumption is justified in our case. The inset shows the tail of the integrated Gaussian distribution (dashed line) compared to a pure exponential decay (solid line) on a logarithmic scale. The difference is relatively small across four orders of magnitude, which shows that the tail of the distribution can plausibly account for the universal behaviour observed in Seebeck coefficient, dc conductivity and torque magnetization measurements. Other distributions result in an even better exponential tail, e.g., the gamma and logistic distributions are asymptotically exponential.

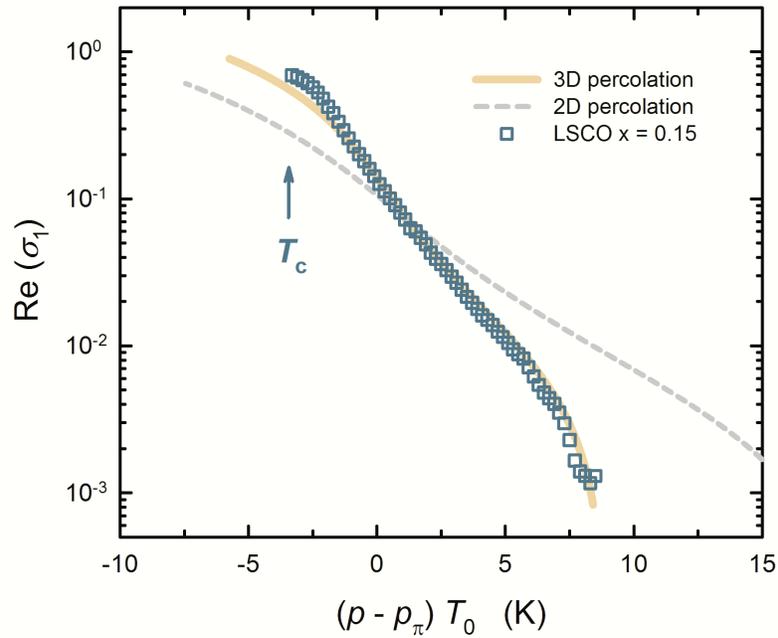

**Fig. S4 |** Comparison between data for LSCO ($x$ = 0.15) and the results of 2D (dashed line, with $p_\pi$ = 0.6) and 3D (full line, with $p_\pi$ = 0.3) site percolation calculations. The same value $T_0$ = 28 K (obtained from nonlinear conductivity) was used in both calculations. This comparison demonstrated that the data are compatible with 3D percolation.

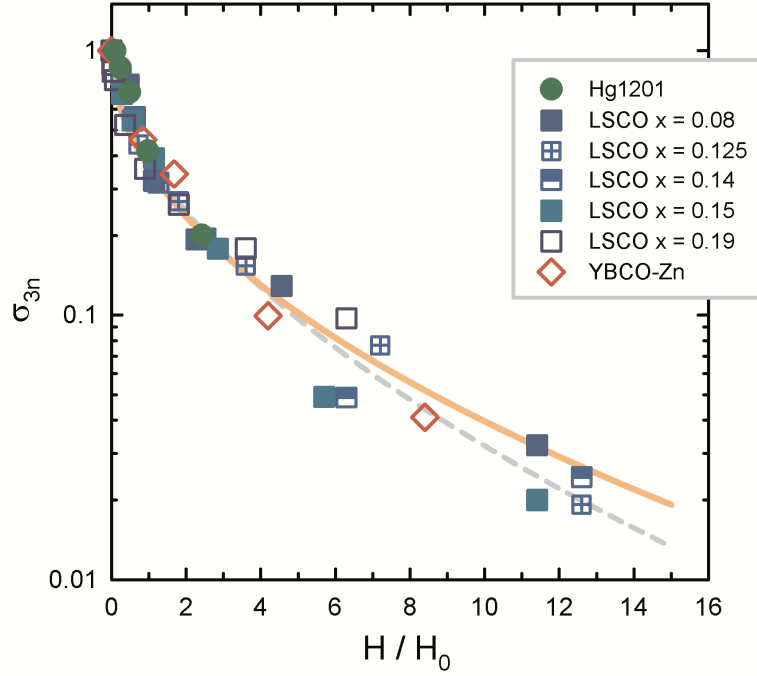

**Fig. S5 |** Magnetic field dependence of the nonlinear conductivity peak (same data as Fig. 2b in the main text) with two different fit results. The solid line is the stretched exponential function with exponent ½, and the dashed line a best fit stretched exponential with exponent 0.59.

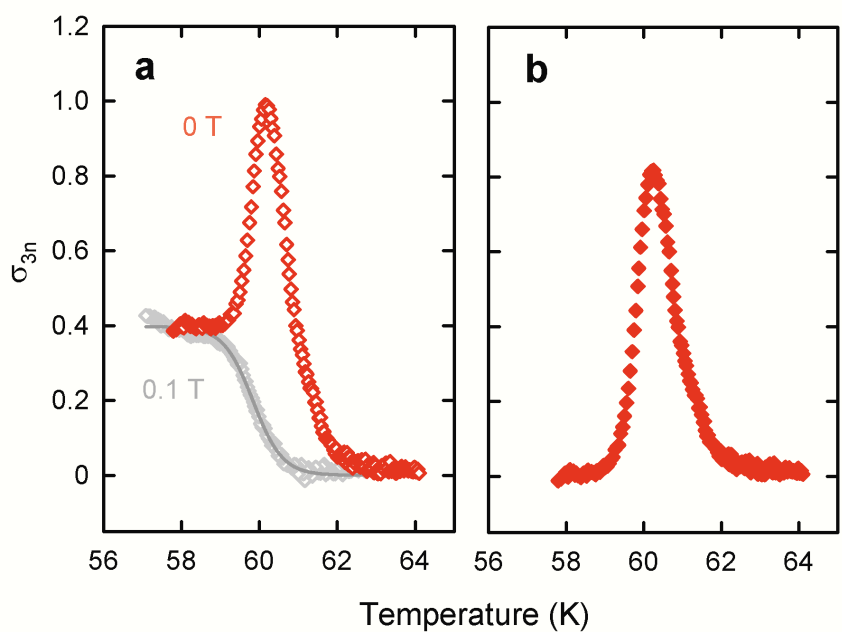

**Fig. S6 | Step-like response subtraction for YBCO-Zn. a,** Raw data in zero field and 0.1 T (where the peak is completely suppressed), with the fitted sigmoidal function (grey line). **b,** Zero-field data with 0.1 T data subtracted.

## Supplementary references

1. Chéenne, N., Mishonov, T. & Indekeu, J., Observation of a sharp lambda peak in the third harmonic voltage response of high-$T_c$ superconductor thin films. *Eur. Phys. J. B* **32,** 437 (2003).

2. Lavrov, A. N., Tsukada, I. & Ando, Y., Normal-state conductivity in underdoped La$_{2-x}$Sr$_x$CuO$_4$ thin films:  Search for nonlinear effects related to collective stripe motion. *Phys. Rev. B* **68,** 094506 (2003).

3. Leviev, G. I., Papikyan, R. S. & Trunin, R., Nonlinear microwave response of YBaCuO in a critical state. *J. Exp. Theor. Phys.* **99,** 359-362 (1990).

4. Bolginov, V. V., Genkin, V. M., Leviev, G. I. & Ovchinnikova, L. V., Nonlinear microwave response of YBCO single crystal in constant magnetic field. *J. Exp. Theor. Phys.* **88,** 1229-1235 (1999).

5. Gallitto, A. A. & Vigni, M. L., Harmonic emission at microwave frequencies in YBa$_2$Cu$_3$O$_7$ single crystals near $T_c$. *Physica C* **305,** 75-84 (1999).

6. Coffey, M. W., Coupled nonlinear electrodynamics of type-II superconductors in the mixed state. *Phys. Rev. B* **46,** 567(R) (1992).

7. Coffey, M. W. & Clem, J. R., Unified theory of effects of vortex pinning and flux creep upon the rf surface impedance of type-II superconductors. *Phys. Rev. Lett.* **67,** 386 (1991).

8. Barišić, N. et al., Demonstrating the model nature of the high temperature superconductor HgBa$_2$CuO$_{4+\delta}$. *Phys. Rev. B* **78,** 054518 (2008).

9. Peligrad, D.-N., Mehring, M. & Dulčić, A., Short-wavelength cutoff effects in the ac fluctuation conductivity of superconductors. *Phys. Rev. B* **67,** 174515 (2003).

10. Grbić, M. S. et al., Microwave measurements of the in-plane and c-axis conductivity in HgBa$_2$CuO$_{4+\delta}$: Discriminating between superconducting fluctuations and pseudogap effects. *Phys. Rev. B* **80,** 094511 (2009).

11. Grbić, M. S. et al., Temperature range of superconducting fluctuations above $T_c$ in YBa$_2$Cu$_3$O$_{7-\delta}$ single crystals. *Phys. Rev. B* **83,** 144508 (2011).

12. Hopfengärtner, R., Hensel, B. & Saemann-Ischenko, G., Analysis of the fluctuation-induced excess dc conductivity of epitaxial YBa$_2$Cu$_3$O$_7$ films: Influence of a short-wavelength cutoff in the fluctuation spectrum. *Phys. Rev. B* **44,** 741 (1991).

13. Dorsey, A. T., Linear and nonlinear conductivity of a superconductor near $T_c$. *Phys. Rev. B* **43,** 7575 (1991).

14. Tranquada, J. M. et al., Evidence of unusual superconducting correlations coexisting with stripe order in $La_{1.875}Ba_{0.125}CuO_4$. *Phys. Rev. B* **78,** 174529 (2008).

15. Chang, J. et al., Decrease of upper critical field with underdoping in cuprate superconductors. *Nature Phys.* **8,** 751-756 (2012).

16. Muniz, R. A. & Martin, I., Method for detecting superconducting stripes in high-temperature superconductors based on nonlinear resistivity measurements. *Phys. Rev. Lett.* **107,** 127001 (2011).

17. Stauffer, D. & Aharony, A., *Introduction to Percolation Theory.* (Taylor & Francis, London, 1994).

18. Boyer, M. C. et al., Imaging the two gaps of the high-temperature superconductor $Bi_2Sr_2CuO_{6+x}$. *Nature Phys.* **3,** 802-806 (2007).

19. Lv, Y.-F. et al., Mapping the electronic structure of each ingredient oxide layer of high-$T_c$ cuprate superconductor $Bi_2Sr_2CaCu_2O_{8+\delta}$. *Phys. Rev. Lett.* **115**, 237002 (2015).

20. Kato, T., Maruyama, T., Okitsu, S. & Sakata, H., Doping dependence of two energy scales in the tunneling spectra of superconducting $La_{2-x}Sr_xCuO_4$. *J. Phys. Soc. Jpn.* **77**, 054710 (2008).

21. Kirkpatrick, S., Percolation and conduction. *Rev. Mod. Phys.* **45**, 574 (1973).

22. Nakamura, M., Conductivity for the site-percolation problem by an improved effective-medium theory. *Phys. Rev. B* **29**, 3691 (1984).

23. Tinkham, M., *Introduction to Superconductivity.* (McGraw-Hill, New York, 1996).

24. Reber, T. J. et al., Prepairing and the filling gap in the cuprates from the tomographic density of states. *Phys. Rev. B* **87,** 060506(R) (2013).

25. Lang, G. et al., Spatial competition of the ground states in 1111 pnictides. *Phys. Rev. B* **94,** 014514 (2016).

26. Salamon, M. B. & Jaime, M., The physics of manganites: Structure and transport. *Rev. Mod. Phys.* **73,** 583 (2001).

27. Phillips, J. C., Saxena, A. & Bishop, A. R., Pseudogaps, dopants, and strong disorder in cuprate high-temperature superconductors. *Rep. Prog. Phys.* **66,** 2111-2182 (2003).

28. Singer, P. W., Hunt, A. W. & Imai, T., $^{63}$Cu NQR evidence for spatial variation of hole concentration in La$_{2-x}$Sr$_x$CuO$_4$. *Phys. Rev. Lett.* **88,** 047602 (2002).

29. Bobroff, J. et al., Absence of static phase separation in the high-$T_c$ cuprate YBa$_2$Cu$_3$O$_{6+y}$. *Phys. Rev. Lett.* **89,** 157002 (2002).

30. Rybicki, D. et al., $^{63}$Cu and $^{199}$Hg NMR study of HgBa$_2$CuO$_{4+\delta}$ single crystals. *arxiv*:1208.4690 (2012).

31. Mihajlović, D., Kabanov, V. V. & Müller, K. A., The attainable superconducting $T_c$ in a model of phase coherence by percolating. *Europhys. Lett.* **57,** 254-259 (2002).

32. Abrikosov, A. A., Possible explanation of the pseudogap in high-temperature cuprates. *Phys. Rev. B* **63,** 134518 (2001).

33. Mosqueira, J., Cabo, L. & Vidal, F., Structural and $T_c$ inhomogeneities inherent to doping in La$_{2-x}$Sr$_x$CuO$_4$ superconductors and their effects on the precursor diamagnetism. *Phys. Rev. B* **80,** 214527 (2009).

34. Phillips, J. C., Percolative theories of strongly disordered ceramic high-temperature superconductors. *Proc. Nat. Acad. Sci. USA* **107,** 1307-1310 (2010).

35. Ovchinnikov, Y. N., Wolf, S. A. & Kresin, V. Z., Intrinsic inhomogeneities in superconductors and the pseudogap phenomenon. *Phys. Rev. B* **63,** 064524 (2000).

36. Gomes, K. K. et al., Visualizing pair formation on the atomic scale in the high-T$_c$ superconductor Bi$_2$Sr$_2$CaCu$_2$O$_{8+d}$. *Nature* **447**, 569 (2007).

37. Honma, T. & Hor, P. H. Quantitative connection between the nanoscale electronic inhomogeneity and the pseudogap of Bi$_2$Sr$_2$CaCu$_2$O$_{8+\delta}$ superconductors. *Physica C* **509,** 11-15 (2015).

38. Barišić, N. et al., Hidden Fermi-liquid behavior throughout the phase diagram of the cuprates. http://arxiv.org/abs/1507.07885 (2015).

39. Grissonnanche, G. et al., Direct measurement of the upper critical field in cuprate superconductors. *Nature Comm.* **5,** 3280 (2014).

40. Sonier, J. E. et al., Hole-doping dependence of the magnetic penetration depth and vortex core size in YBa$_2$Cu$_3$O$_y$: Evidence for stripe correlations near 1/8 hole doping. *Phys. Rev. B* **76,** 134518 (2007).